\documentclass[pra,aps,showpacs,twocolumn,superscriptaddress]{revtex4-1}

\usepackage{amsfonts}
\usepackage{amsmath}
\usepackage{amssymb}
\usepackage{graphicx}

\begin{document}

\title{Quantum tomography in position and momentum space}

\date{\today}

\author{J. Casanova}
\affiliation{Departamento de Qu\'{\i}mica F\'{\i}sica, Universidad del Pa\'{\i}s Vasco UPV/EHU, Apartado 644, 48080 Bilbao, Spain}
\author{ C. E. L\'{o}pez}
\affiliation{Departamento de F\'{\i}sica, Universidad de Santiago de Chile, USACH, Casilla 307 Correo 2 Santiago, Chile}
\author{J. J. Garc\'ia-Ripoll}		
\affiliation{Instituto de F\'isica Fundamental, CSIC, Serrano 113-bis, 28006 Madrid, Spain}
\author{C. F. Roos}
\affiliation{Institut f\"ur Quantenoptik und Quanteninformation, \"Osterreichische Akademie der Wissenschaften, Otto-Hittmair-Platz 1, A-6020 Innsbruck, Austria}
\affiliation{Institut f\"ur Experimentalphysik, Universit\"at Innsbruck, Technikerstrasse 25, A-6020 Innsbruck, Austria}
\author{E. Solano}
\affiliation{Departamento de Qu\'{\i}mica F\'{\i}sica, Universidad del Pa\'{\i}s Vasco UPV/EHU, Apartado 644, 48080 Bilbao, Spain}
\affiliation{IKERBASQUE, Basque Foundation for Science, Alameda Urquijo 36, 48011 Bilbao, Spain}


\begin{abstract}
We introduce a method of quantum tomography for a continuous variable system in position and momentum space. We consider a single two-level probe interacting with a quantum harmonic oscillator by means of a class of Hamiltonians, linear in position and momentum variables, during a tunable time span. We study two cases: the reconstruction of the wavefunctions of pure states and the direct measurement of the density matrix of mixed states. We show that our method can be applied to several physical systems  where high quantum control can be experimentally achieved.

\end{abstract}

\maketitle

\section{Introduction}

The tomography of low-dimensional systems~\cite{paris04} is a well established process in fundamentals of quantum physics and quantum information~\cite{poyatoscirac}. Measuring a finite set of observables on identical copies of the system, one is capable of reconstructing the density matrix of a quantum state with a finite precision. However, practical experiments have a limited accuracy and the number of measurements grows exponentially with the size of the Hilbert space. Furthermore, such schemes do not have a direct translation to setups where the degrees of freedom are continuous, as is the case of atomic beams~\cite{mcalister94, mlyneck97}, harmonic oscillators in trapped ions~\cite{monroe96,leibfried96,reviewions,reviewblatt,wineland09,brown11,harlander11}, cavity QED~\cite{reviewharoche,walther06,brune08}, circuit QED~\cite{wallraff04,blais06,houck07,martinis08,deppe08,fink09,martinis09,Astafiev10,naturecircuitqed},  nanomechanical resonators~\cite{nanobose}, and superconducting qubits~\cite{moooij99,martinis02,vion06,martinis10}.

In the case of continuous variable systems, the measurement of an infinite number of observables would be required and, in consequence, other simplifications or schemes have to be developed~\cite{reviewraymer,sascha95}. In linear optics or in atomic ensembles, where interactions are typically quadratic, one assumes that the generated states are Gaussian. That is, they are fully characterized by first and second field quadrature moments, the so called covariance matrix~\cite{reviewhayashi}. In this case, again, a finite set of measurements suffices to determine the state, its entanglement properties, and all other observables.

However, not all physically realizable states are Gaussian. For example, photon-added states and superpositions of Fock states~\cite{martinis09} are interesting resources for quantum information processing. For a continuous variable description of such non-Gaussian states the most popular solution is the reconstruction of the Wigner function~\cite{davidovich97}, a quasi-probability distribution that contains the same information as the density matrix. 

In this paper, we propose a method for reconstructing the state of a continuous variable system in position and momentum space, be for a pure state or a mixed state. To accomplish this purpose, we couple the quantum system to a two-level probe with an adequate interaction. For other reconstruction methods using different interactions and experimental requirements, see, for example, Ref.~\cite{wineland96}. We show that, by monitoring the probe, we can obtain enough information to reconstruct the wave function of the system in position and momentum space.

This paper is organized as follows. In Sec. II, we present the model Hamiltonian and we develop the method to obtain the wave function when the system is in a pure state and a mixed state. In Sec. III, we propose a physical implementation of our method. In Sec IV, we present our concluding remarks.

\section{Quantum tomography of continuous-variable systems}

From an operational point of view, our proposal requires the generic interaction Hamiltonian between a two-level probe and a $\ell$-dimensional system
\begin{equation}
\label{interaction}
H=\hbar g \sigma_{\vec{n}} \otimes \left( \vec{\alpha} \cdot \vec{R} + \vec{\beta} \cdot \vec{P}  \right).
\end{equation}
Here, $\sigma_{\vec{n}}$ is a probe Pauli operator along ${\vec{n}}$, $\vec{\alpha}$ and $\vec{\beta}\in\mathbb{R}^{\ell}$ contain control parameters, while $\vec{R} = (X_1,\ldots,X_{\ell})$ and $\vec{P} = (P_1,\ldots,P_{\ell})$ are the dimensionless position and momentum operators of the system. When the probe is coupled to the system, they exchange quantum information and, by monitoring the state of the probe as a function of time, we will be able to reconstruct the wave function $\psi_s(\vec{R})$ of the oscillator system. 

In the following we will consider the reconstruction of a quantum state, be it in a pure or a mixed state. For simplicity and without loss of generality, we will develop the theory for one-dimensional systems, $\ell=1$ in Eq.~(\ref{interaction}) and present useful examples.

\subsection{Pure states}

In many experimental setups suitable for quantum tomography, the measurable quantity is the probe observable $\sigma_z$; this could be for example a trapped and laser-manipulated two-level atom or a superconducting qubit. In this sense, the value of $| \psi_{s} (x) |^2$ will be obtained by monitoring the evolution of $\langle \sigma_{z} \rangle$~\cite{naturedirac}. To achieve this, we set the control parameters such that Eq.~(\ref{interaction}) turns into $H= \hbar g \alpha  \sigma_x X$, where $\sigma_x$ is the Pauli matrix along $x$ direction and $X = (a+a^\dagger)/\sqrt{2}$ is the dimensionless position quadrature. The evolution of the two-level probe can be calculated as $\langle \sigma_{z} \rangle_{t}= \langle \Psi | U^{\dag} \sigma_{z}  U |\Psi \rangle$, where $U=\exp{(-i H t/\hbar)}$ and $| \Psi \rangle$ is the total probe-system wavefunction. It is convenient to write
\begin{equation}
\langle \sigma_{z} \rangle_{t}=\langle \Psi | \cos{(k X)} \sigma_{z}|\Psi \rangle + \langle \Psi | \sin{(k X)} \sigma_{y}|\Psi \rangle,
\label{sigmaz1}
\end{equation}
with $k=2g\alpha t$. From this expression, we are able to obtain $|\psi_{s} (x)|^2$ by considering two different initial states for the probe, leading to two sets of measurements. In the first one, we consider the probe decoupled from the system such that we have an initial state $|\Psi \rangle = |\psi_s \rangle \otimes |\uparrow\rangle_{z}$, such that, $\sigma_{z} |\uparrow\rangle_{z}=|\uparrow\rangle_{z}$. In this case, the expectation value of $\sigma_z$, $P_{z}^{\rm e}(k)$, will be
\begin{equation}
P_{z}^{\rm e}(k)=
\int_{-\infty}^{\infty}\cos{(kx)} |\psi_{s}(x)|^2 \ dx .
\label{Pzeven}
\end{equation}
We observe that the quantity $P_{z}^{\rm e}(k)$ is related to the even part  of $|\psi_{s}(x)|^2$. Then, the second set of measurements provides us the corresponding odd part. This is found by preparing system initially in the state $|\Psi \rangle = |\psi_s \rangle \otimes |\uparrow\rangle_{y}$ such that $\sigma_{y} |\uparrow \rangle_{y}=|\uparrow\rangle_{y}$.  In this case, the expectation value of $\sigma_z$ is given by
\begin{equation}
P_{z}^{\rm o}(k)=\int_{-\infty}^{\infty}\sin{(kx)} |\psi_{s} (x)|^2 \ dx \label{Pzodd}.
\end{equation}

\begin{figure*}[t]
\centering{
\includegraphics[width=55.5mm]{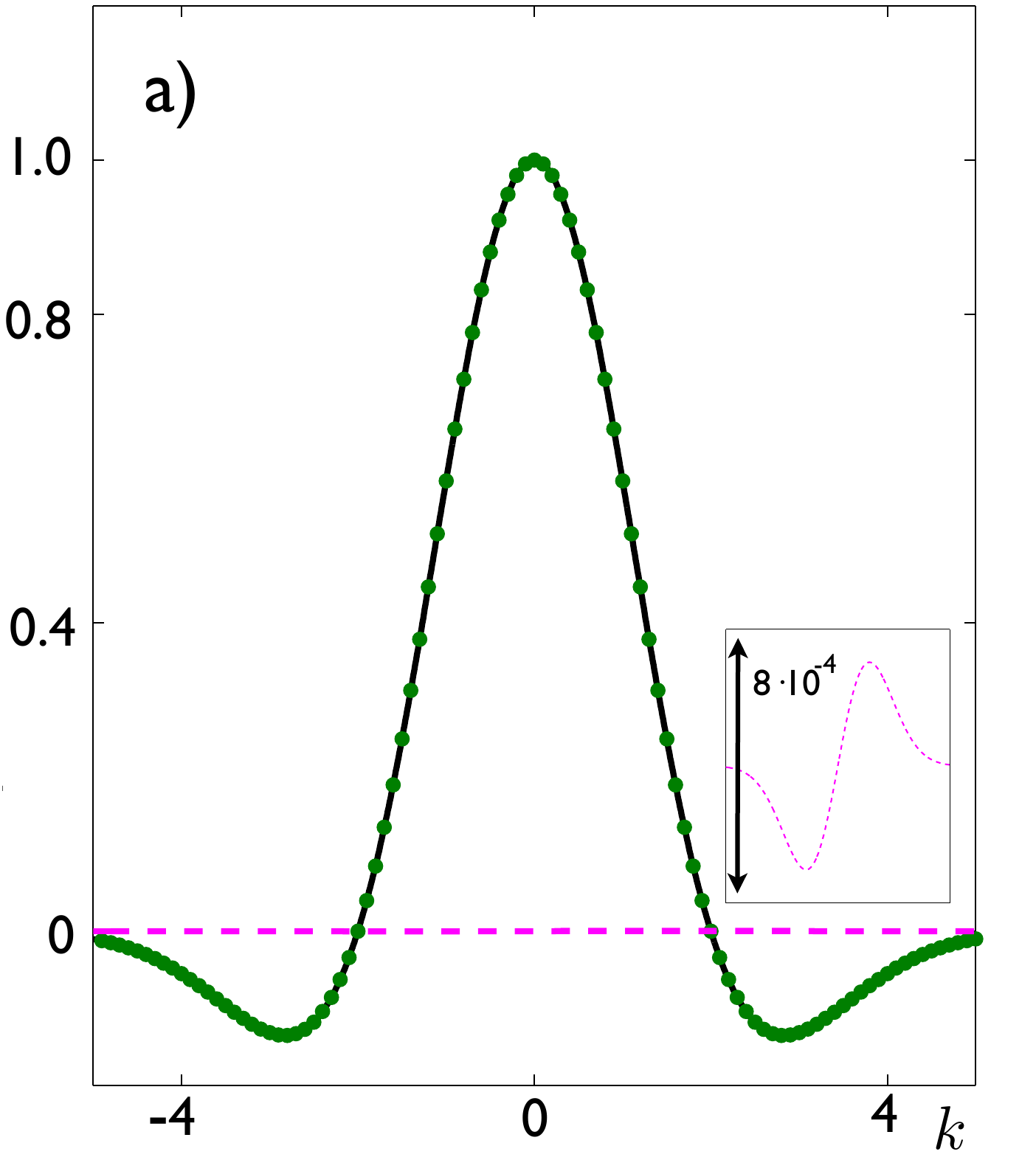}
\includegraphics[width=57mm]{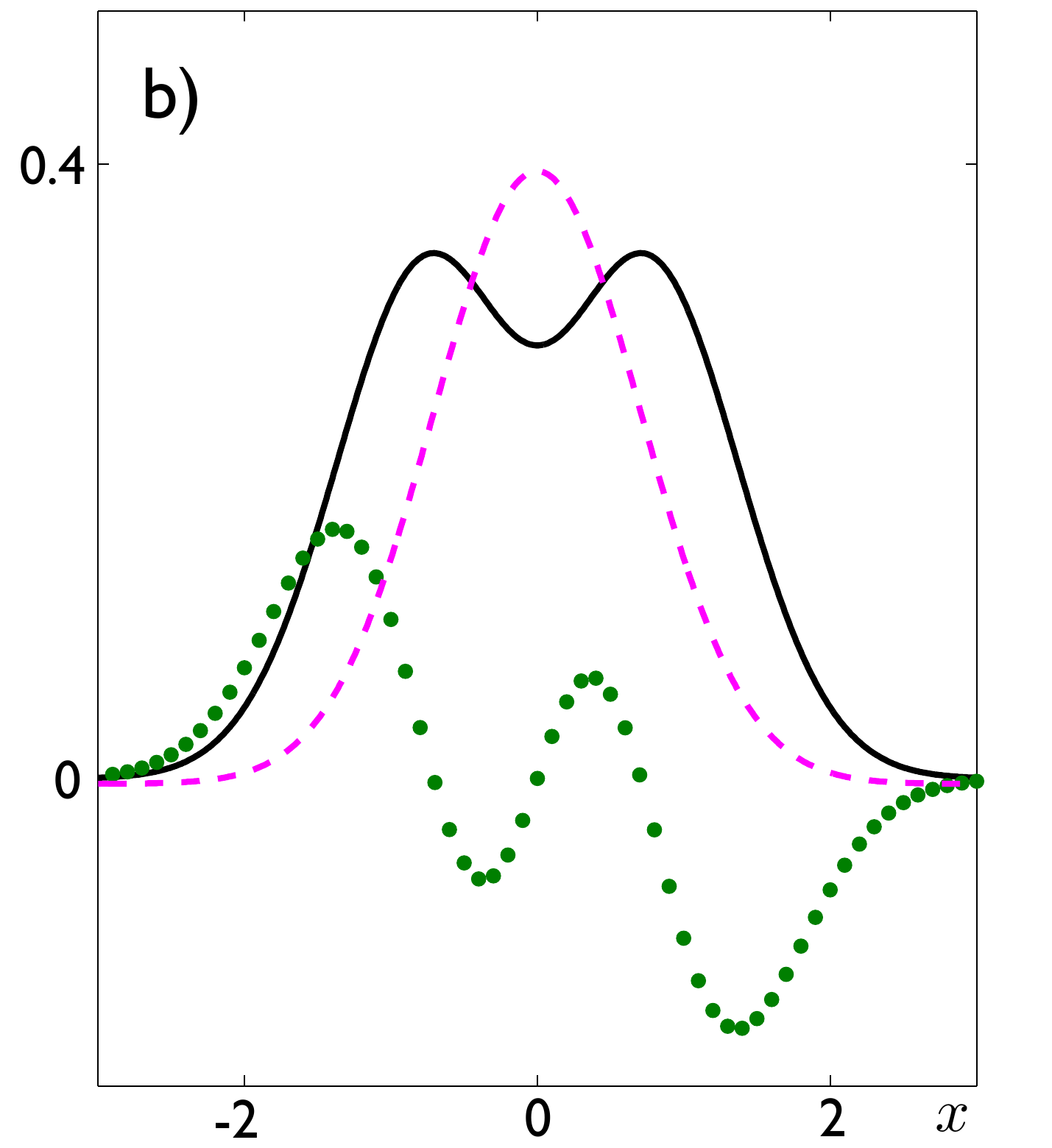}
\includegraphics[width=59mm]{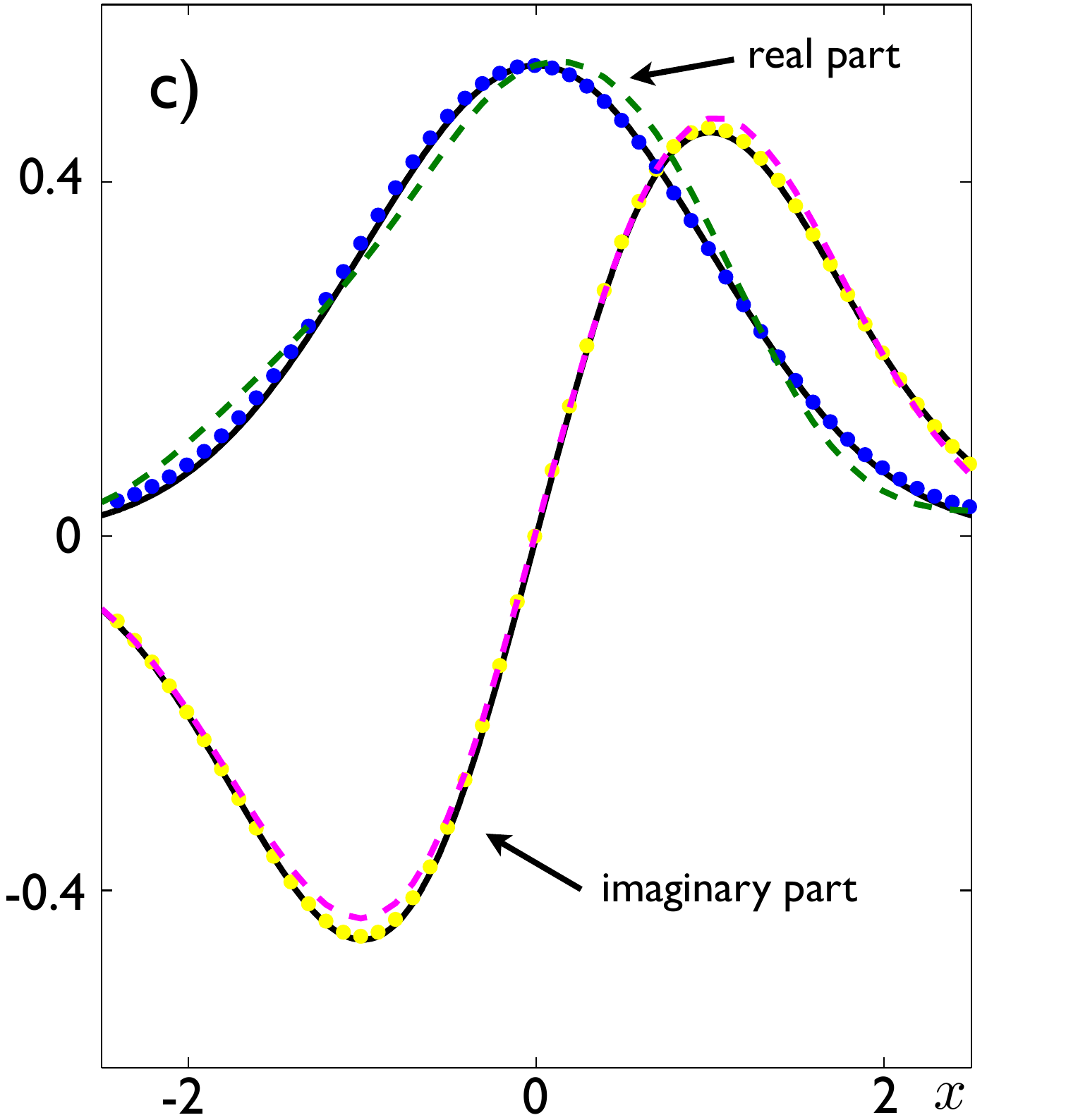}}
\caption{(Color online) Different steps needed to solve the wave function in position space of the  state $|\psi_{s}\rangle=(|0\rangle+ i |1\rangle)/\sqrt{2}$ using the method for pure states. a) Plots of $P_{z}^{\rm e}(k)+i P_{z}^{\rm o}(k)$ , which is real in our example (solid-black line), and the real and imaginary parts of $ \tilde{P}_{z}^{\rm e}(k)+i \tilde{P}_{z}^{\rm o}(k)$ (dotted-green and dashed-magenta lines). The inset represents the blow-up of the corresponding imaginary part, $\tilde{P}_{z}^{\rm o}(k)$. b) Plots of  $|\psi_{s}(x)|^2$, (solid-black line),  $\mathrm{Re} \left[G(x)\right] $ (dotted-green line) and $\mathrm{Im} \left[G(x)\right] $ (dashed-magenta line).  c) Comparison between the real and imaginary parts of the wave function obtained using the method, (dotted blue and yellow lines) and the exact values calculated with the analytical expressions for $\langle x|\psi_s \rangle= (\langle x|0\rangle+ i \langle x|1\rangle)/\sqrt{2}$, (solid-black lines), in this case the relation between $\alpha$ and $\beta$ is $\beta/\alpha = 0.5\times10^{-3} $. The dashed green and magenta lines in (c) correspond to the results of the real and imaginary parts of $\psi$  calculated now with a less strict choice of $\beta/\alpha = 0.1$, which shows the minor effects of the second order contributions of $q$ in Eq~(\ref{aproximacion}) and, henceforth, the robustness of our method. }
\label{fig1}
\end{figure*} 

Combining the results of the two sets of measurements, we obtain the squared absolute value of the wave function by noticing that
\begin{equation}
|\psi_{s}(x)|^2=  {\cal F}^{-1}[P_{z}^{\rm e}(k)+i P_{z}^{\rm o}(k)] ,
\label{psis}
\end{equation}
where ${\cal F}^{-1}[f(k)] = \frac{1}{2\pi}\int_{-\infty}^{\infty} f(k) e^{-ikx} dk $ is the inverse Fourier transform of $f(k)$. In this manner, we have shown that the squared absolute value of the system wavefunction can be obtained by monitoring the populations of a two-level probe. When the state of the system has no imaginary components, the measurement of $|\psi_{s}(x)|^2$ is enough to obtain the wave function $\psi_{s}(x)$.  This method for obtaining the wave function can be extended to the case of multimode quantum systems.

To reconstruct the complete wavefunction when the state of the system has complex components, we will need additional information. In this case, we must properly set the control parameters to produce the Hamiltonian
\begin{equation}\label{hampart2}
 H =\hbar g \sigma_x (  \alpha X +  \beta P), 
\end{equation}
with tunable $\alpha$ and $\beta$, and where $P = i(a^{\dag}-  a )/\sqrt{2}$ is the dimensionless momentum quadrature. Similar to the previous case, we can obtain
\begin{eqnarray} 
\tilde{P}_{z}^{\rm e}&=&\langle \psi_{s} |  \cos(k X+q P ) |\psi_{s}\rangle ,\label {Pzeven2} \\
\tilde{P}_{z}^{\rm o}&=&\langle \psi_{s} |  \sin(k X+q P ) |\psi_{s}\rangle,  \label {Pzodd2}
\label{}
\end{eqnarray}
where $k=2gt \alpha$ and $q=2gt \beta$.
Using the Baker-Campbell-Haussdorf formulas, we can write	
\begin{eqnarray}
  \tilde{P}_z^{\rm e} &=& \frac{1}{2}\langle \psi_{s} | e^{i k X+iq P } + \mathrm{H.c.} |\psi_{s}\rangle \\
  &=& \mathrm{Re}\langle \psi_{s} | e^{i k X} e^{iq P } e^{i kq /2} |\psi_{s}\rangle, \nonumber \\
  \tilde{P}_z^{\rm o} &=& \mathrm{Im}\langle \psi_{s} | e^{i k X} e^{iq P } e^{i kq/2} |\psi_{s}\rangle.
\end{eqnarray}
We consider now $\langle qP\rangle \ll 1$ and expand the previous expressions for $\tilde{P}_{z}^{\rm e}$ and $\tilde{P}_{z}^{\rm o}$, involving terms up to first order in~$q P$. Combining these results, we have
\begin{eqnarray}\label{aproximacion}
&&(\tilde{P}_{z}^{\rm e}+i \tilde{P}_{z}^{\rm o})e^{-\frac{i}{2} kq}\simeq\\
&&\quad\quad\int_{-\infty}^{\infty}\psi_{s}^{*}(x)e^{i k x }\left[
1 + i q P + {\cal O}(q^2)\right] \psi_{s} (x) \ d x \, . \nonumber
\end{eqnarray}
In the position representation, $P=-i\partial/\partial_{x}$, we derive 
\begin{eqnarray}\label{G}
&&G(x):=\psi_{s}^{*} (x) \partial_{x}\psi_{s} (x)=\\
&&\quad   {\cal F}^{-1}\left[ \frac{1}{q}(\tilde{P}_{z}^{\rm e}(k)+i \tilde{P}_{z}^{\rm o}(k))e^{-\frac{i}{2} kq} - \frac{1}{q}(P_{z}^{\rm e}(k)+i P_{z}^{\rm o}(k))\right] . \nonumber
\end{eqnarray}
where $q=(\beta / \alpha) k$.

This expression can be combined  with the previous result, where we found $|\psi_s (x)|^2$, producing the ratio
\begin{equation}
\frac{\psi_{s}^{*} (x) \partial_{x}\psi_{s} (x)}{\psi_{s}^{*}(x)\psi_{s}(x)}=\frac{\partial_{x}\psi_{s} (x)}{\psi_{s}(x)}=\frac{G(x)}{|\psi_{s}(x)|^2} \, ,
\label{difeq}
\end{equation}
which results in a differential equation for the wave function in the position space. 
In the general case in which $\psi_{s}(x)$ is a complex function, the knowledge of $f(x) =G(x)/|\psi_{s}(x)|^2  $ allows us to write a set of coupled differential equations
\begin{eqnarray}\label{set}
\frac{\partial \psi_{s}^{\rm r}(x)}{\partial x} &=& f^{\rm r}(x) \psi_{s}^{\rm r}(x) - f^{\rm im}(x) \psi_{s}^{\rm im}(x) \nonumber  \\ 
\frac{\partial \psi_{s}^{\rm im}(x)}{\partial x} &=& f^{\rm r}(x) \psi_{s}^{\rm im}(x) + f^{\rm im}(x) \psi_{s}^{\rm r}(x) ,
\end{eqnarray}
where $\psi_s^{\rm r, im}(x)$ and $f^{\rm  r,  im}(x)$ are the real and imaginary parts of $\psi_{s}(x)$ and $f(x)$. The set of Eqs.~(\ref{set}) can be easily integrated with any numerical protocol given a known initial value, coming from the previous knowledge of  $|\psi_{s}(x) |^2$, and the addition of an irrelevant global phase.

On the other hand, following a similar procedure, the wavefunction in the momentum space can be measured by setting properly the control parameters to exchange the roles of $X$ and $P$. Tuning the control parameters to produce the Hamiltonian $H=\ \hbar g \beta \sigma_x   P$, we can obtain an expression similar to Eq.~(\ref{psis}) but in momentum representation, $|\psi_{s}(p)|^2$. To find $\psi_{s}(p)$, we consider the Hamiltonian~(\ref{hampart2}) but now with the condition $\langle k X \rangle \ll 1$, and we find a differential Eq. for $\psi_{s}(p)$ analogous to Eq.~(\ref{difeq}). We present a pedagogical example in Fig.~\ref{fig1}. In this manner, we have reconstructed the wave function of an arbitrary pure quantum state in the position representation by monitoring the evolution of an observable of the two-level probe.

\begin{figure*}[t]
\centering{
\includegraphics[width=59mm]{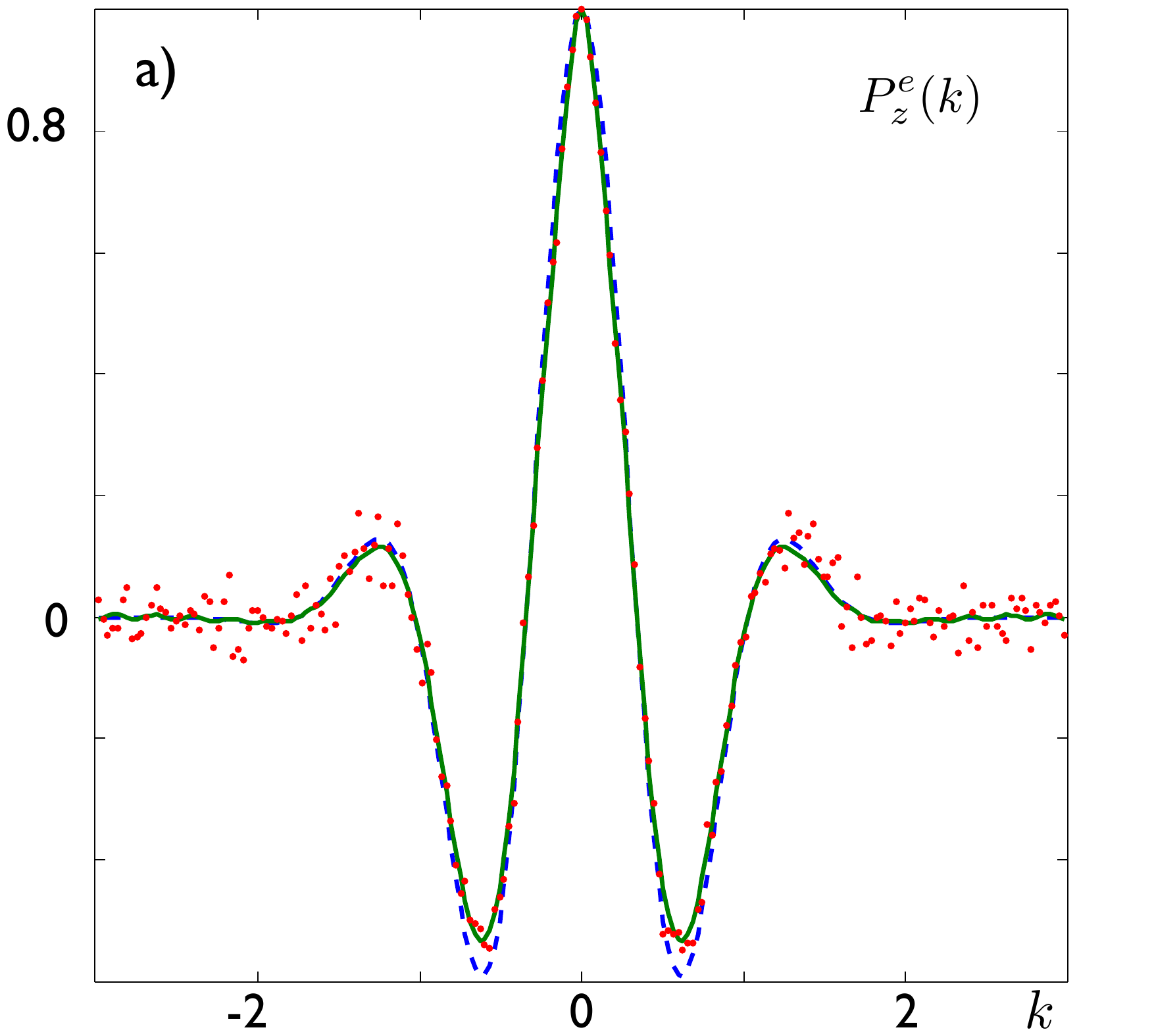}
\includegraphics[width=58mm]{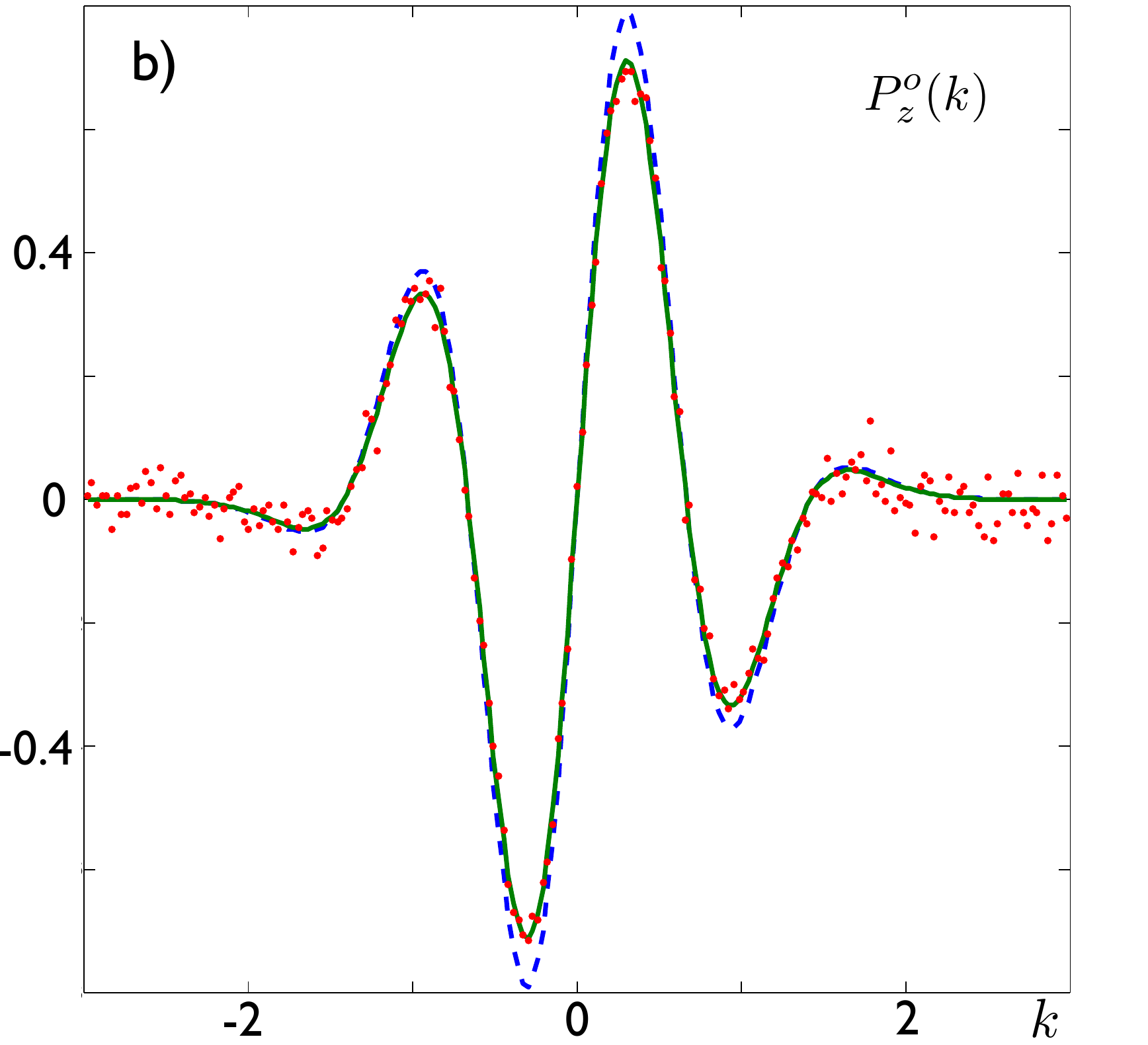}
\includegraphics[width=57mm]{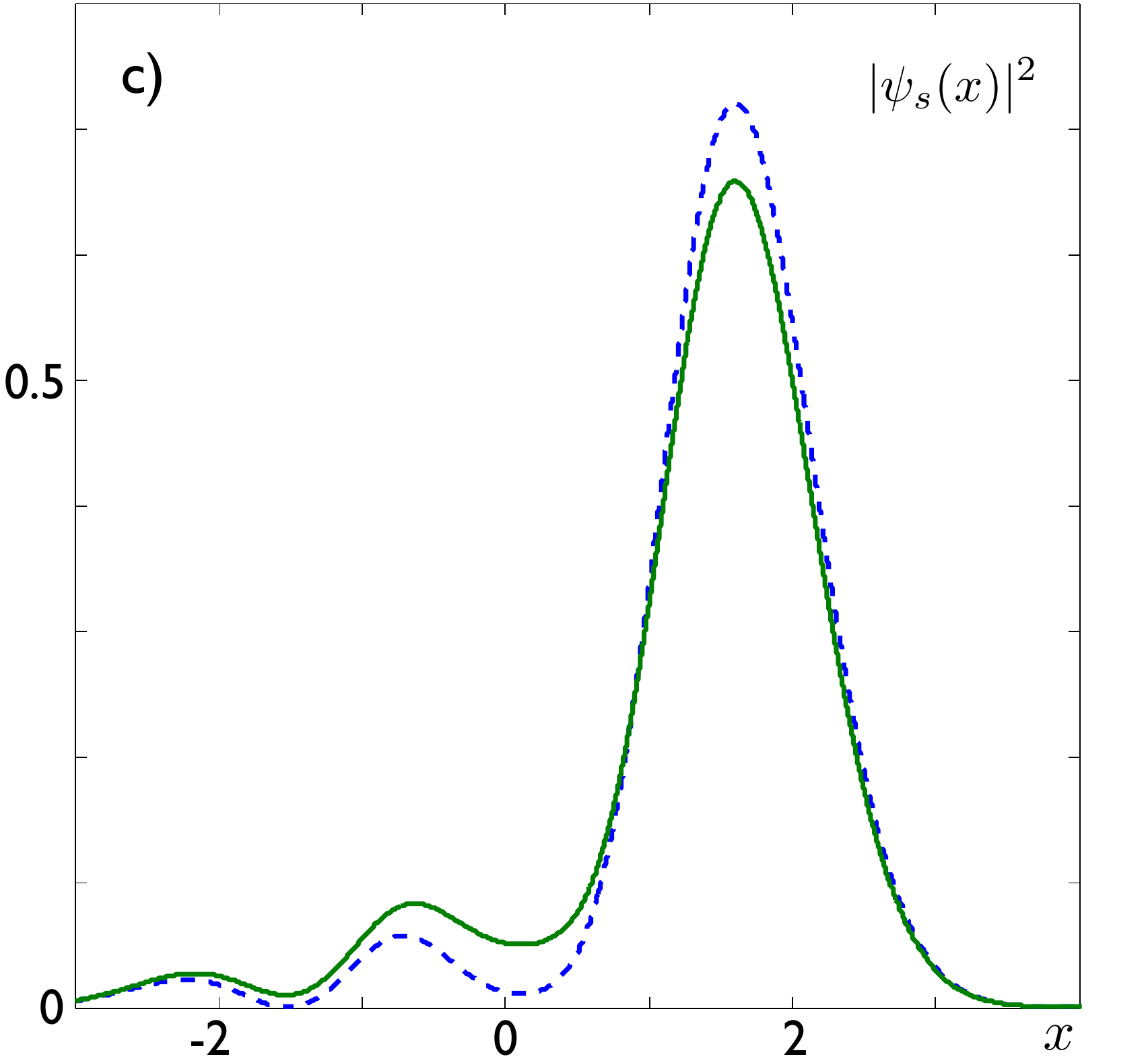}}
\caption{(Color online) a) and b): Plots obtained using the described method for a state $|\psi_{s}\rangle=(|0\rangle+|1\rangle+|2\rangle+|3\rangle)/2$ in a depolarizing channel $\mathcal{E}(|\psi_{s}\rangle \langle \psi_{s}|)=(1-\epsilon)|\psi_{s}\rangle \langle \psi_{s}|+\epsilon I /d $ with $\epsilon=0.1$. Dots show data from numerically simulated experiment. Green line is the fitting curve for the simulated data. Blue line is the expected result when $\epsilon=0$. In c) the squared "wave function" obtained from data in a) and b) (green line) is compared with the expected $|\psi_{s}(x)|^2$ for $\epsilon=0$ (blue dashed line).} \label{fig2}
\end{figure*}

\subsection{Mixed states}

In the laboratory, pure states can only be approximate and, in general, we deal with mixed states. In this case, a quantum system is described by a density matrix instead of a wavefunction. To have a deeper insight, in Fig.~\ref{fig2} we simulate numerically an experiment considering the Hamiltonian $H=\hbar g \alpha \sigma_{x} X$. Note that in both examples, see Figs.~\ref{fig1} and \ref{fig2}, the dimensionless parameters $k$ and $X$ are considered in a range where typical experimental times are several orders of magnitude smaller than the involved decoherence ones~\cite{davidovich97,naturedirac}. We compare then the wavefunction of a pure state with a similar one under the effects of a depolarizing channel $\mathcal{E}(|\psi_{s}\rangle \langle \psi_{s}|)=(1-\epsilon)|\psi_{s}\rangle \langle \psi_{s}|+\epsilon I /d $. This channel maps the pure density matrix $|\psi_{s}\rangle \langle \psi_{s}|$ onto a superposition of this pure state and a fully mixed state $I/d$ with probability $\epsilon$, being $d$ the dimension of the system. In Fig.~\ref{fig2}, we show the robustness of the method for reconstructing pure states in the presence of some decoherence processes. This numerical simulation was implemented by sorting a random number and comparing it with the theoretical value of occupation probabiliy $P_{z}^{\rm e}(k)$, assuming $1\% $ of error in detectors. If the random number is less than the corresponding $P_{z}^{\rm e}(k)$, we assume the probe to be in the excited state, otherwise it would be in the ground state. Without the added error, we would approach the theoretical values for a large sampling. In the following, we extend our method to measure density matrices.

Let us now consider the general case of the system described by the density matrix in the Fock basis $\rho_{s} = \sum c_{n,m}|n\rangle\langle m|$. We will show that all matrix elements, $c_{n, m}$, can be obtained using the same technique considered in the pure case. The choice of $\beta = 0$ in the one-dimensional version of the Hamiltonian~(\ref{interaction}) determines  the evolution of the expectation value of the observable $\sigma_z$,  given by
\begin{equation}
\langle \sigma_{z} \rangle_t   = {\rm Tr}[\rho_s \otimes |\Phi\rangle \langle \Phi | e^{i g t \sigma_{x} X} \sigma_{z} e^{-i g t \sigma_{x} X}].
\end{equation}
Performing both sets of measurements for the two initial probe states, we find
\begin{equation}\label{result1}
P_{z}^{\rm e}(k)+i P_{z}^{\rm o}(k) = \sum_{n,m} c_{n,m} \int dx \ e^{ik x} \psi_{m}^{*}(x)\psi_{n}(x),
\end{equation}
where $k=2g t$ and $\psi_{j}(x) = \langle x|j\rangle$ is the wavefunction associated to the $j$-th Fock state. This last expression yields a set of equations for the coefficients $c_{n,m}$. However, since the wavefunctions of the Fock states are real, Eq.~(\ref{result1}) only provides partial knowledge about the density matrix. In consequence, we need to introduce more relations to complete a set of linearly independent equations allowing us to find all $c_{n,m}$. To this end, we let the system evolve under the free-energy Hamiltonian $H=\hbar \omega a^{\dag}  a$ for a time $t=t_{0}$. This evolution produces a change in the matrix elements of the system, i.e., $\rho_{s}(t_{0})= \exp{(-i\omega a^{\dag}a t_{0})} \rho_s \exp{(i\omega a^{\dag}a t_{0})} = \sum_{n,m} c_{n,m}\exp{(-i(n-m)\omega t_{0})} |n\rangle\langle m|$. Under this new initial probe state, Eq.~(\ref{result1}) becomes
\begin{eqnarray}
P_{z}^{\rm e}(k)+i P_{z}^{\rm o}(k)  &=& \sum_{n,m} c_{n,m}e^{-i(n-m)\omega t_{0}} \notag \\
&&\times \int dx \ e^{ik x} \psi_{m}^{*}(x)\psi_{n}(x). \label{result2}
\end{eqnarray}
Choosing properly different values of the parameter $t_{0}$, depending on the size of the density matrix $\rho_{s}$, we obtain a set of linear equations allowing us the estimation of all matrix elements $c_{n,m}$.

The proposed method for measuring the density matrix of a harmonic oscillator, via direct estimation of its matrix elements from Eq.~(\ref{result2}), might be compared to some early Wigner function reconstructions~\cite{Vogel89,Smithey93}. In those works, marginal distributions of the Wigner function, associated to arbitrary field quadratures, are obtained using homodyne detection techniques. In our case, we encode the information of the density matrix elements in the level statistics of a probe that is coupled to the system with a Hamiltonian proportional to $X$. The additional phase rotation of the initial state, via the use of Hamiltonian $H=\hbar \omega a^{\dag}  a$, allows us to access information related to other quadratures, that is, linear superpositions of $X$ and $P$. However, our method relies on probe measurements at arbitrary interaction times, making it difficult to trace any linear dependence with the involved Hamiltonians along the quantum evolution.

\section{Physical Implementation}

The methods described above could be implemented in different physical setups where two-level probes interact with quantum harmonic oscillators, as is the case of cavity QED and trapped ions in quantum optics, or circuit QED in mesoscopic physics. The case of trapped ions can be easily implemented combining red and blue sideband excitations with suitable phases~\cite{quantumwalkion,kleintheory,kleinexperiment}. Here, we will give an example of how to construct the required family of Hamiltonians in the case of cavity QED. We consider then a field mode oscillator with energy $\hbar \omega$ and a two-level probe atom with transition energy $\hbar \omega_0 $. Let us assume that the system and the probe are in the joint state $|\Psi\rangle=|\Phi \rangle_p  \otimes |\psi_{s} \rangle $, where $|\Phi\rangle_p$ denotes the state of the probe and $|\psi_{s} \rangle$ denotes the state of the system to be measured. We allow them to interact via a Jaynes-Cummings coupling while the probe is driven by a phase-sensitive coherent field with frequency $\omega_L$. The total Hamiltonian describing this situation reads~\cite{solano03}
\begin{eqnarray}\label{hamiltonian1}		
H&=&\hbar \omega_0 |e\rangle \langle e |+\hbar \omega a^{\dag}  a +\hbar g ( \sigma^{\dagger} a + \sigma a^{\dagger}) \notag \\
&&+\hbar\Omega  (\sigma^{\dagger} e^{-i\omega_L t+i \varphi}+\sigma e^{i\omega_L t-i \varphi}),
\end{eqnarray}
where $a^{\dag} (a)$ is the creation (annihilation) operator of the oscillator mode and $g$ is the coupling strength between the probe and the system. In the strong-driving regime, $\Omega \gg \{ g, \varphi \}$, and under resonant conditions, $\omega_0 = \omega = \omega_L$, the Hamiltonian in the interaction picture reads
\begin{eqnarray}
 H_{\rm int} & = & \frac{\hbar g}{2}  \big( |+_{\varphi} \rangle \langle +_{\varphi}|- |-_{\varphi}\rangle \langle -_{\varphi}| \big) ( a e^{- i \varphi} +a^{\dag} e^{i \varphi} )\notag \\
&\equiv & \hbar g \, \sigma_x ^{\varphi} X_{\varphi} \equiv \hbar g \, \sigma_x ^{\varphi} ( \alpha X + \beta P ), \label{hamvarphi}
\end{eqnarray}

\vspace*{0cm}

where $\sigma_x ^{\varphi} = |+_{\varphi} \rangle \langle +_{\varphi}|- |-_{\varphi}\rangle \langle -_{\varphi}|$, in the dressed basis $|\pm_{\varphi}\rangle = (| g \rangle \pm e^{i\varphi} |e\rangle)/\sqrt 2$, and $X_{\varphi}=(  a e^{-i \varphi}+ a^{\dag} e^{i \varphi})/2$. Therefore, we have shown how to build the desired Hamiltonian, similar to Eq.~(\ref{interaction}).

\section{Summary}

In summary, we have developed a method for measuring the  wavefunction of a quantum system in a pure state in position and momentum space, or the density matrix of a mixed state. This is achieved by suitable monitoring the evolution of a two-level probe coupled to the system via a Hamiltonian linear in position and momentum variables. We expect that the proposed methods contribute to the already mature field of quantum state tomography.

\vspace*{0.7cm}

\acknowledgements
We acknowledge support from Basque Government BFI08.211 and IT472-10; Fondecyt 1121034; PBCT-CONICYT PSD54; Financiamiento Basal para Centros Cient\'{\i}ficos y Tecnol\'{o}gicos de Excelencia; MICINN FIS2009-10061 and FIS2009-12773-C02-01; UPV/EHU UFI 11/55; QUITEMAD; SOLID, CCQED, and PROMISCE European projects.

\end{document}